\documentclass[%
 reprint,
superscriptaddress,
showpacs,preprintnumbers,
 amsmath,amssymb,
 aps,
]{revtex4-1}

\usepackage[utf8]{inputenc}
\usepackage[english]{babel}
\usepackage{amssymb}
\usepackage{bm}
\usepackage[dvipsnames]{xcolor}
\usepackage{bbm}
\usepackage{bbold}

\usepackage{latexsym}
\usepackage{amsmath}
\usepackage{float}

\usepackage{soul}
\usepackage{ulem}
\usepackage{graphicx}
\usepackage{dcolumn}
\usepackage{bm}
\usepackage{color}
\usepackage{natbib}
\usepackage{hyperref}
\usepackage{cleveref}
\begin{document}
\date{\today}

\title{Tuning the electronic structure and magnetic coupling in armchair B$_2$S nanoribbons using strain and staggered sublattice potential  }
\author{Moslem Zare }
\email{mzare@yu.ac.ir}
\affiliation{Department of Physics, Yasouj University, Yasouj, Iran 75914-353, Iran}

\date{\today}

\begin{abstract}
Magnetically-doped two dimensional honeycomb lattices are potential candidates for application in future spintronic devices.
Monolayer B$_2$S has been recently unveiled as a desirable honeycomb monolayer with an anisotropic Dirac cone.
Here, we investigate the carrier-mediated exchange coupling, known as Ruderman-Kittel-Kasuya-Yoshida (RKKY) interaction, between two magnetic impurity moments in armchair-terminated B$_2$S nanoribbons in the presence of strain and staggered sublattice potential.
By using an accurate tight-binding model for the band structure of B$_2$S nanoribbons near the main energy gap, we firstly study the electronic properties of all infinite-length armchair B$_2$S nanoribbons (ABSNRs), with different edges, in the presence of both strain and staggered potential.

It is found that the ABSNRs show different electronic and magnetic behaviors due to different edge morphologies.
The band gap energy of ABSNRs depends strongly upon the applied staggered potential $\Delta$ and thus one can engineer the electronic properties of the ABSNRs via tuning the external staggered potential.
A complete and fully reversible semiconductor (or insulator) to metal transition has been observed via tuning the external staggered potential, which can be easily realized experimentally.
A prominent feature is the presence of a quasiflat edge mode, isolated from the bulk modes in the ABSNRs belong to the family $M=6p$, with $M$ the width of the ABSNR and $p$ an integer number.
As a key feature, the position of the quasi-flatbands in the energy diagram of ABSNRs can be shifted by applying the in-plane strains $\varepsilon_x$ and $\varepsilon_y$. At a critical staggered potential ($\Delta_c\sim 0.5$ eV), for nanoribbons of arbitrary width, the quasi-flatband changes to a perfect flatband.

The RKKY interaction has an oscillating behaviour in terms of the applied staggered potentials, such that for two magnetic adatoms randomly distributed on the surface of an ABSNR the staggered potential can reverse the RKKY from antiferromagnetism to ferromagnetism and vice versa. The RKKY in terms of the width of the ribbon has also an oscillatory behavior.
It is shown that the magnetic interactions between adsorbed magnetic impurities in ABSNRs can be manipulated by careful engineering of external staggered potential.
Our findings pave the way for applications in spintronics and pseudospin electronics devices based on ABSNRs.

\end{abstract}

\maketitle

\section{Introduction}

The intriguing prospect of the potential nanoelectronic and optoelectronic applications, which may take advantage of the novel two dimensional (2D) materials with exciting electronic properties, has inspired researchers to explore possibilities of such materials with outstanding characteristics.
As a typical two-dimensional material, pristine graphene is one of the most attractive materials due to its outstanding potential applications in many fields~\cite{Novoselov2005}, but unfortunately, the lack of a finite band gap in graphene is a major obstacle for using graphene in nanoelectronic and optoelectronic devices. However, a big challenge for graphene science is how to open a substantial band gap for graphene without significantly degrading its excellent outstanding advantages in graphene based nanoelectronic devices~\cite{X.LiScience,Sonprl}.
In the aspect of nanoelectronic and optoelectronic 2D research, the major issue is the availability of 2D materials with a wide band gap window in their band structure.
In this regard, B$_2$S monolayer, an atomically thin layer of boron and sulfur atoms arranged in a honeycomb pattern with perfect planar structure, appears in the research field again, by using global structure search method and first principles calculation combined with tight-binding model~\cite{Yu.Zhao,P.Li}. It is reported that this new 2D anisotropic Dirac cone material has a Fermi-velocity up to $10^6 m/s$ in the same order of magnitude as that of graphene. It is thermally and dynamically stable at room temperature and is a potential candidate for future nanoelectronic applications~\cite{Yu.Zhao,P.Li}. B$_2$S monolayer is the first pristine honeycomb lattice with a tilted anisotropic Dirac cone structure, stabilized by sulfur atoms in free standing condition. Since boron atom has one electron less than carbon, all the reported 2D boron-based Dirac cone materials have much more complicated geometries in comparison with the pristine honeycomb structure of graphene~\cite{Gonzalez2007nrl,Ciuparu2004,F.Liu2010,An2006,Eremets2001}.

In the last decades, magnetic atoms embedded in a non-magnetic host material have been intensively studied in solid state physics due to their functionalities for application in spintronic devices and magnetic recording media~\cite{pesin,Dietl2000}.
Dilute magnetic semiconductors, as potential materials for spintronics and optoelectronics, have been studied since early 90-es. These investigations resulted in establishing a unified picture of the nature of indirect exchange interaction between magnetic adatoms, known as the  Ruderman-Kittel-Kasuya-Yosida (RKKY) interaction~\cite{Ruderman, Kasuya, Yosida}, mediated by a background of conduction electrons of the host material.
In diluted metals and semiconductors it is often the dominating magnetic interaction and has played a key part in the development of magnetic phases, e.g., spiral~\cite{moslem-si, Mahroo-RKKY}, spin-glass~\cite{pesin, Eggenkamp, Liu87},
ferromagnetic (FM)~\cite{Vozmediano, Brey, Priour, Matsukura, Ko, Ohno-science} and antiferromagnetic (AFM)~\cite{Minamitani, Loss15}.
This long-range spin-spin interaction leads to spin relevant effects in giant magnetoresistance devices~\cite{Baibich88,Binasch89}, spin filters~\cite{F.Ye-EPL}, drives ferromagnetism in heavy rare-earth elements ~\cite{I.D.Hughes} as well as in diluted magnetic semiconductors ~\cite{Dietl2000}.
It was shown that the RKKY interaction consists of three terms, namely Heisenberg, Ising, and Dzyaloshinskii–Moriya (DM) on the surface of zigzag silicene nanoribbons as well as the three dimensional topological insulators~\cite{pesin,moslem-si,JJZhu}, and the competition between them leads to rich spin configurations.
An additional term, namely spin-frustrated has discovered in a three-dimensional Weyl semimetal (WSM) that along with the Dzyaloshinskii-Moriya term lies in the plane perpendicular to the line connecting two Weyl points~\cite{M.V.Hosseini}.

In a spin polarized system~\cite{fariborz-sp} and a material with multi-band structure~\cite{fariborz-mos2}, these oscillations become more complicated than a monotonic oscillation with $\sin(2k_{\rm F}R)$ behavior, where $k_{\rm F}$ is the wave vector of the electrons (holes) at the Fermi level and $R$ is the distance of two magnetic impurities. In addition, it is important to note that the magnitude of the RKKY interaction can be severely affected by the density of states (DOS) at the Fermi level~\cite{fariborz-blg, Mahroo-RKKY}. Owing to the bipartite nature of the honeycomb sublattice, the RKKY coupling in graphene is highly sensitive to the direction of the distance vector between impurities~\cite{sherafati-g, fariborz-blg}. In materials with spin-orbit interaction of Rashba type~\cite{Mahroo-single}, the exchange interaction depends on the direction of the magnetic moments and, as a result, the RKKY interaction becomes anisotropic~\cite{Mahroo-RKKY}.

We have recently addressed the problem of isolated magnetic adatoms placed on silicene~\cite{unpublished} and phosphorene~\cite{MoslemBP1} sheets as well as on zigzag silicene nanoribbons~\cite{moslem-si} and bilayer phosphorene nanoribbons~\cite{MoslemBP2}.
In a detailed study, we found that the RKKY interaction in silicene can be written in an anisotropic Heisenberg form for the intrinsic
case where the spin coupling could realize various spin models, e.g., the XXZ-like spin model~\cite{unpublished}.
In another work, it has concluded that the RKKY interaction in the bulk phosphorene monolayer is highly anisotropic and the magnetic ground-state of two magnetic adatoms can be tuned by changing the spatial configuration of impurities as well as the chemical potential varying \cite{MoslemBP1}.
Importantly, the occurrence of these magnetic phases not only depends on the magnetic impurity concentration, but also on the concentration of free carriers in the host material~\cite{Eggenkamp95}.

This effective interaction can also be viewed as an indirect coupling mediated by pure spin current in quantum spin Hall systems, due to the helicity ~\cite{J.Gao2009}. This interaction oscillates as a function of the distance between two magnetic adatoms (with wavelength $\pi/k_F$), due to the sharpness of the Fermi surface. Besides these practical magnetic phases, the RKKY interaction can provide information about the intrinsic properties of the material, since this coupling is proportional to the spin susceptibility of the host system.

In the last decades, dilute magnetic semiconductors have emerged as a research hotspot due to their functionalities for application in spintronic devices and magnetic recording media.
Inducing magnetism in otherwise nonmagnetic 2D materials has been a subject of intense research due to the unique physical characteristics originating from 2D confinement of electrons, foe concurrent applications in electronic and optoelectronic devices~\cite{fabian, Babar,W.Han}.

Controllable magnetic properties of nanoribbon-based spintronic devices allow the development of the next generation of magnetic and spintronic devices to be realized, and thus much attention has been focused on determining the magnetic properties of 2D honeycomb nanoribbons~\cite{Klinovaja13,moslem-si,MoslemBP2,Duan17}.

Motivated by the future potential of the honeycomb nanoribbons decorated by magnetic impurities, in this work we have addressed the problem of indirect exchange coupling between localized magnetic moments mediated by the conduction electrons of B$_2$S nanoribbons with armchair-terminated edges.
Within the tight-binding model we exploit the Green's function formalism (GF), to reveal how the RKKY interaction between two magnetic impurities, placed on a B$_2$S nanoribbon, is affected by mechanical strains in the presence of a sublattice staggering potential.

It is found that armchair B$_2$S nanoribbons (ABSNRs) show different electronic and magnetic behaviors due to different edge morphologies.
The band gap energy of ABSNRs depends strongly upon the applied staggered potential $\Delta$, and thus one can engineer the electronic properties of the ABSNRs with desirable characteristics via tuning the external staggered potential.

A complete and fully reversible semiconductor (or insulator) to metal transition has been observed via tuning the external staggered potential, which can be easily realized experimentally.
Interestingly, for the ABSNRs belong to the family $M=6p$, with $M$ the width of the ABSNR and $p$ an integer number, one can see that a band gap, in which a quasi-flatband completely detached from the bulk bands, is always observed.
As a key feature, the position of the quasi-flatbands in the energy diagram of ABSNRs can be shifted by applying the in-plane strains $\varepsilon_x$ and $\varepsilon_y$.
At a critical staggered potential ($\Delta_c \sim 0.5$ eV), for ABSNRs with any width, the quasi-flatband changes to a perfect flatband.

It is shown that the RKKY interaction has an oscillating behaviour in terms of the applied staggered potentials, such that for two magnetic adatoms randomly distributed on the surface of an ABSNR the staggered potential can reverse the RKKY from antiferromagnetism to ferromagnetism and vice versa.
Meanwhile, the RKKY interaction has an oscillatory behavior in terms of the width of the ribbon.

This paper is organized as follows: In Sec.~\ref{sec:theory}, we introduce the system under consideration, i.e., an armchair-terminated B$_2$S nanoribbon under the influence of strain and staggered potential applied to it. A tight-binding model Hamiltonian for monolayer B$_2$S is presented and then the band spectrum of ABSNRs with different edge configurations, under a staggered potential, have been investigated then, we introduce the theoretical framework which will be used in calculating the RKKY interaction, from the real space Green’s function. After that, we discuss our numerical results for the proposed magnetic doped ABSNRs in the presence of both strain and staggered sublattice potential. Finally, our conclusions are summarized in Sec.~\ref{sec:summary}.

\begin{figure}
\includegraphics[width=8.2cm]{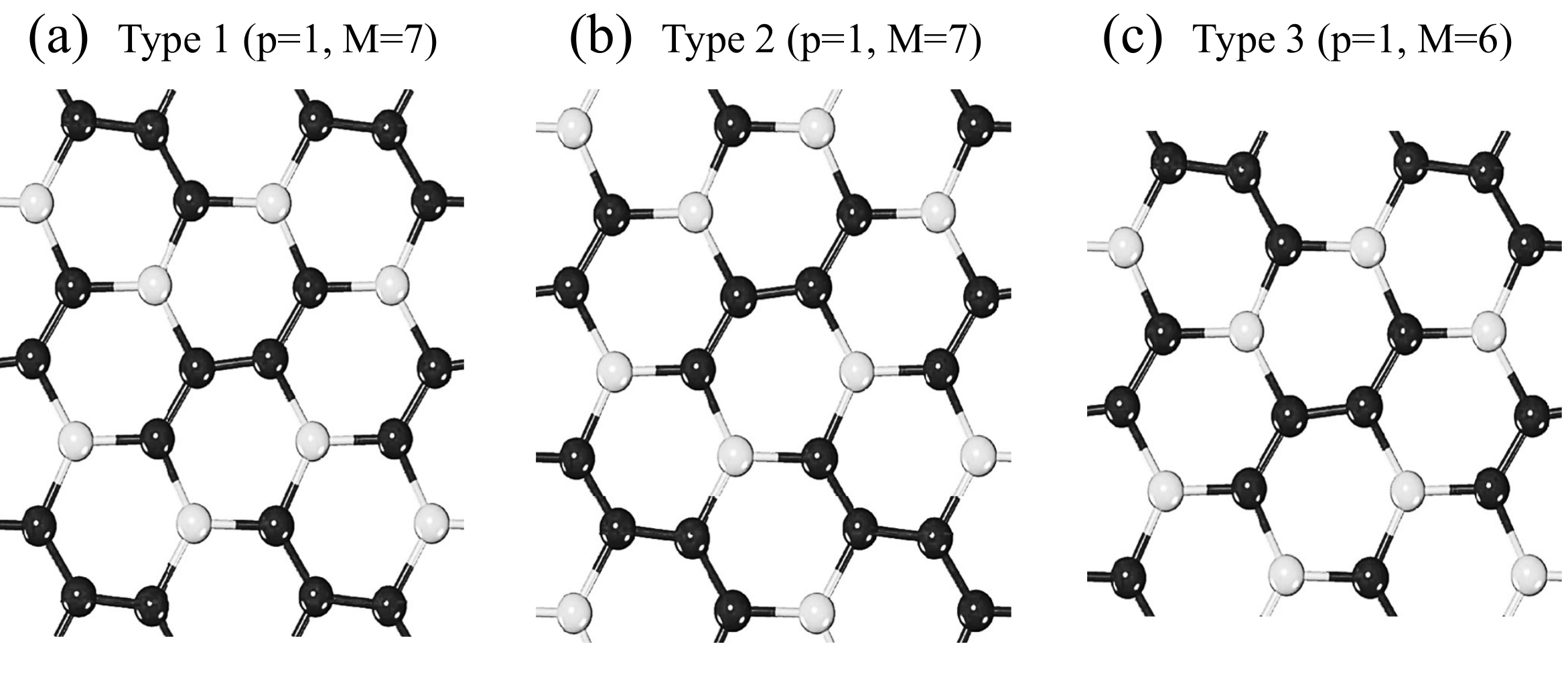}
\includegraphics[width=8.2cm]{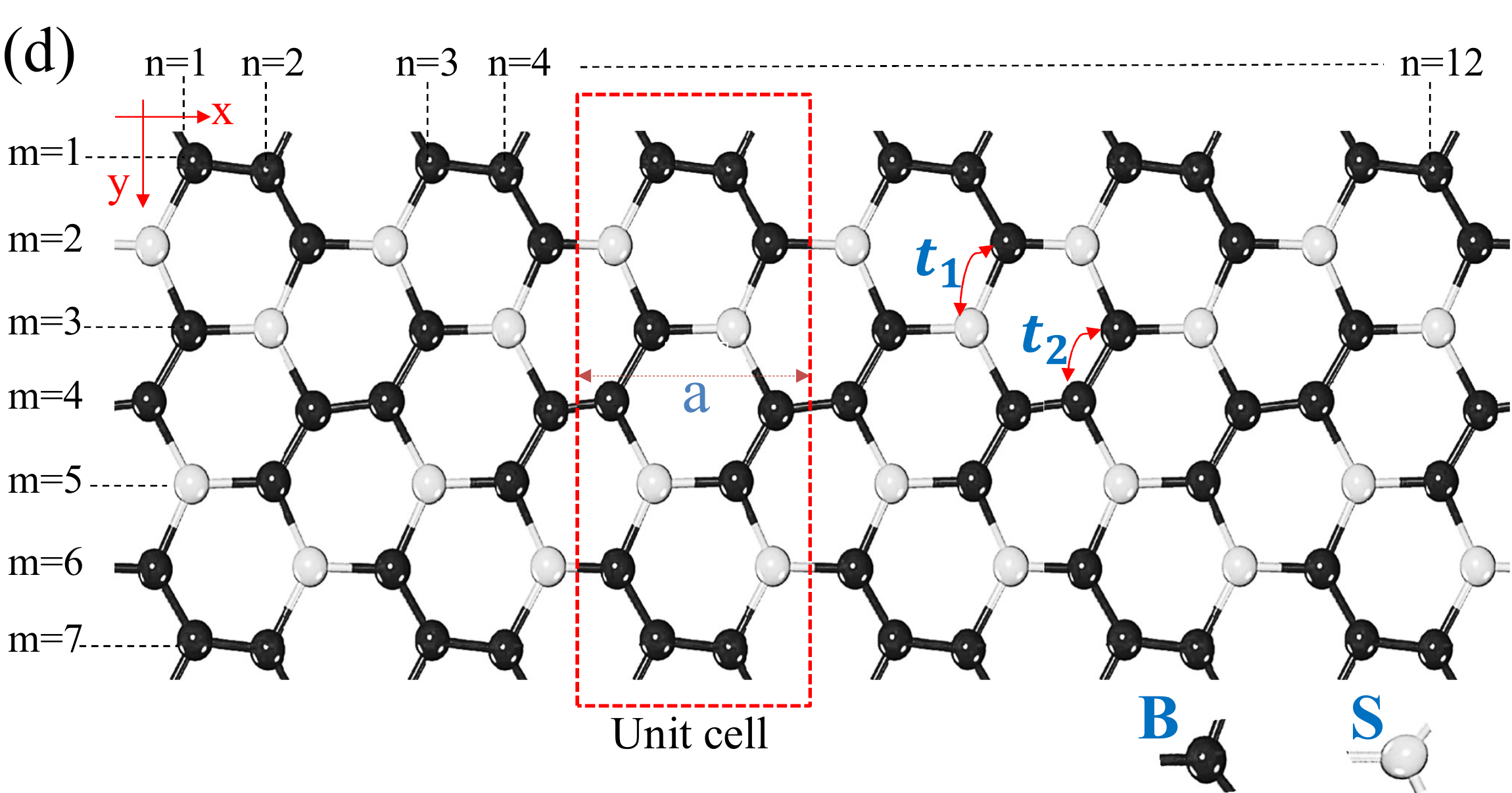}
\caption{(Color online)
Schematic illustration of the optimized geometry structure of armchair B$_2$S nanoribbons (ABSNRs). The top panels : armchair B$_2$S nanoribbons of type 1 (a), Type 2 (b), and Type 3 (c). An example of an ABSNR of type 1 with width $M=7$ and length $N=12$, in which the edges (or the 1D periodicity direction) lie along the $x$ direction, is shown in the bottom panel (d). The red-dashed rectangle represents the super unit cell. The number of atoms in the vertical zigzag lines across the ABSNRs width, $M$ , is used to indicate the width of a B$_2$S nanoribbon and accordingly we use the number of vertical zigzag lines ($N$) to measure its length (in the units of a)
The black balls represent the boron atoms ($B$) and the white ones correspond to the sulfur atoms ($S$). The nearest-neighbor hopping parameters, used in the tight-binding Eq.~\ref{eqn:TBhamiltonian}, are denoted by $t_1$ and $t_2$. For simplicity, each atom is labeled with a set $(n, m)$, where $n$, $m$ represent the $x$ and $y$ coordinates of the lattice sites.}
\label{schem}
\end{figure}

\section{Theory and model}\label{sec:theory}

\subsection{The RKKY interaction}\label{sec:RKKY}
To study indirect magnetic interaction between two local moments in armchair-terminated B$_2$S nanoribbons, we consider the indirect exchange coupling between magnetic impurities to be of the RKKY form, mediated by the conduction electrons in armchair B$_2$S nanoribbon system.
The contact magnetic interaction between the spin of itinerant electrons and two magnetic impurities located at positions ${\bf r}$ and ${\bf r'}$ with magnetic moments ${\bf S}_1$ and ${\bf S}_2$ is given by

\begin{equation}
V = - \lambda \ ( {\bf S}_1 \cdot {\bf s(r)} + {\bf S}_2 \cdot {\bf s(r')} ),
\label{hamilint}
\end{equation}
where ${\bf s(r)}, {\bf s(r')} $ are the conduction electron spin densities at positions ${\bf r}$ and ${\bf r'}$ and $\lambda$ is the contact potential between the impurity spins and the itinerant carriers.

Using a second-order perturbation~\cite{Ruderman,Kasuya,Yosida}, the effective magnetic interaction between local moments induced by the free carrier spin polarization, the RKKY interaction, which arises from quantum effects, reads as

\begin{equation}
E({\bf r},{\bf r'}) = J ({\bf r},{\bf r'})  {\bf S}_1 \cdot {\bf S}_2,
\label{RKKYE}
\end{equation}

The RKKY interaction $J ({\bf r},{\bf r'}) $ is explained using the static carrier susceptibility, the response of the charge density $n({\bf r})$ to a perturbing potential $V({\bf r'})$, {\it i.e.}, $ \chi({\bf r},{\bf r'}) \equiv  \delta n({\bf r}) / \delta V({\bf r'})$ , which is given by
\begin{equation}
J ({\bf r},{\bf r'}) = \frac{\lambda^2 \hbar^2 }{4} \chi ({\bf r},{\bf r'}).
\label{RKKYJ}
\end{equation}

The static spin susceptibility can be written in terms of the integral over the unperturbed Green's function $G^0 $ as
\begin{equation}
\chi ({\bf r},{\bf r'}) =
- \frac{2}{\pi} \int^{\varepsilon_F}_{-\infty} d\varepsilon \
{\rm Im} [G^0 ({\bf r}, {\bf r'}, \varepsilon) G^0 ({\bf r'},{\bf r}, \varepsilon)],
\label{chiGG}
\end{equation}
where $\varepsilon_F$ is the Fermi energy. The expression for the susceptibility may be obtained by using the spectral representation of the Green's function
\begin{equation}
G^0 ({\bf r},{\bf r'},\varepsilon)= \sum_{n,s} \frac{\psi_{n,s}({\bf r})\psi^{*}_{n,s}({\bf r'})}{\varepsilon+i\eta - \varepsilon_{n,s}},
\label{GFspct}
\end{equation}
where $\psi_{n,s}$ is the sublattice component of the unperturbed eigenfunction with the corresponding energy $\varepsilon_{n,s}$. For a crystalline structure, ${n,s}$ denotes the band index and spin. Substituting Eq. (\ref{GFspct}) into Eq. (\ref{chiGG}), after integration over energy, we will get the result for the RKKY interaction.
The analytical background of this approach has been already described in details in previous works~\cite{moslem-si,MoslemBP2} and is not rediscussed here. Finally, from those analytical calculations the RKKY interaction can be expressed in the following desired result

\begin{eqnarray}
\chi({\bf r},{\bf r'}) &&=2 \sum_{\substack{n,,s \\ n',s'}}[ \frac{f(\varepsilon_{n,s})-f(\varepsilon_{n',s'})}{\varepsilon_{n,s}-\varepsilon_{n',s'}}\nonumber\\
&&\times \psi_{n,s}({\bf r})\psi^{* }_{n,s}({\bf r'})\psi_{n',s'}({\bf r'})\psi^{*}_{{ n'}s'}({\bf r})].
\label{chiE}
\end{eqnarray}
where, $f(\varepsilon)$, is the Fermi function.
This is a well-known formula in the linear response theory that is the main equation in this work.

\subsection{Armchair B$_2$S nanoribbons}\label{subsec:armchair}

In this section, we reintroduce B$_2$S monolayer to the layered-material family as an anisotropic material for optoelectronic and spintronic applications.
Since indirect exchange interaction between magnetic moments is significantly affected by the electronic structure of the host material, tailoring electronic properties of this nanostructure is crucial. To do so, the electronic structures of armchair-terminated B$_2$S nanoribbons are studied using tight-binding model.

As very recently reported~\cite{Yu.Zhao}, the most energetically stable structure of B$_2$S monolayer predicted by using global structure search method and first principles calculation combined with tight-binding model, is shown in Fig.\ref{schem} from which we can see that the planar 2D structure consists of honeycomb lattices, similar to graphene. This honeycomb structure is a global minimum in the space of all possible 2D arrangements of B$_2$S in which each hexagonal ring is distorted with the bond angles ranging from $114$ \AA~ to $123$ \AA, because B and S atoms have different covalent radii and electronegativities~\cite{Yu.Zhao,P.Li}.

From the Figs.\ref{schem} (a-c), it has been demonstrated that different types of ribbons are specified by their edge geometry and width. As seen, the armchair B$_2$S nanoribbons can be divided into three families, {\it i.e.}, $M=6p+1$ (type 1, with $N$ boron atoms on the edges of the ABSNR), $M=6p+1$ (type 2, with $N/2$ boron atoms on the edges of the ABSNR), $M=6p$ (type 3, with $N$ boron atoms on the one edge and $N/2$ boron atoms on the another edge of the ABSNR), with $p$ as an integer number.
The bottom panel shows an ABSNR of type 1 in which the armchair edge is at the $x$ direction. For an infinite-sized ABSNR this system shows translational symmetry along $x$ axis. The red-dashed rectangle represents the super unit cell.
The usage of such geometry division had two aims, first to evaluate the behaviour of the infinite length ABSNRs, and second to investigate some important finite size effects which will be discussed further.

The geometrical structure of the pristine armchair edge B$_2$S nanoribbon, lying in the $xy$ plane, is depicted in Fig.\ref{schem} (d). As shown in this figure, each hexagon consists of four $B$ atoms and two $S$ atoms, with an orthogonal primitive cell with a space group of $PBAM$ and a point group $D_2h$.
As shown in Fig.\ref{schem}, the bonding length between two adjacent $B$ atoms ($B$-$B$ bonds) was calculated to be $1.62$ \AA, from the relaxed structure whereas the distances between $B$ and $S$ atoms ($B$-$S$ bonds), are all of the same length $1.82$ \AA~\cite{Yu.Zhao}.
The black balls correspond to the boron atoms ($B$) and the white ones correspond to the sulfur atom ($S$). For simplicity, each atom is labeled with a set $(n, m)$, where $n$, $m$ represent the $x$ and $y$ coordinates of the lattice sites.
The number of atoms in the vertical zigzag lines across the ABSNRs width, $M$, is used to indicate the width of a BSNR and accordingly we use the number of vertical zigzag lines ($N$) to measure its length.

From the analysis of symmetry and orbital characters of the wave functions in a B$_2$S monolayer it is clear that a tight-binding (TB)
model involving just the tilted $p_z$ orbitals should be able to describe the band structure of this 2D layered material near the Fermi level~\cite{Yu.Zhao}. We begin with describing this nearest-neighbor effective tight-binding Hamiltonian, given by

\begin{equation}\label{eqn:TBhamiltonian}
H = - \sum_{\langle i,j \rangle} t_{ i,j } c_{i}^{\dagger} c_{j}  + \sum_i U_{i,B} c_{i,B}^{\dagger} c_{i,B}  +
\sum_i U_{i,S} c_{i,S}^{\dagger} c_{i,S},
\end{equation}

with nearest neighbor hopping energies $t_{1}=0.8$ eV and $t_{2}=1.7$ eV ~\cite{Yu.Zhao} (see Fig.~\ref{schem}(d), the bottom panel) where $c_{i}^{(\dag)}$ is the annihilation (creation) operator of the electron at the $i$-th lattice site and $\sum_{\langle i,j \rangle}$ sums over all nearest neighbor pairs.
We consider a general situation where a staggered sublattice potential is applied throughout the sheet, $\Delta/2$ for sublattices $B$, and $-\Delta/2$  for sublattices $S$.
However, it has been found that the onsite energies for $B$ and $S$ atoms are $V_{S}=5.4$ eV and $V_{S}=6.4$ eV, respectively and thus, the corresponding parameters $U_{i,S(B)}$, are as follows: $U_{i,B}=V_{B}+\Delta/2$ for sublattices $B$, and $U_{i,S}=V_{S}-\Delta/2$ for sublattices $S$, respectively.

Having an accurate tight-binding model, as presented in the equation above (Eq.\ref{eqn:TBhamiltonian}), we can numerically calculate the momentum space dispersion of a monolayer armchair nanoribbon of B$_2$S. To do so, we make a one-dimensional (1D) Fourier transform (owing to the translational invariant along the ribbon direction, $x$), in accordance with Bloch’s theorem obtained from $\sum_{{\bm k}}\psi_{{\bm k}}^{\dag} H_k \psi_{{\bm k}}$, with respect to the $x$ direction:

\begin{equation}\label{eq:H-k1}
 H_k=H_{00}+H_{01}e^{-i k_x
a}+H_{01}^\dagger e^{i k_x a}
\end{equation}
in which $a$ is the unit-cell length along the x-axis. Moreover, $H_{00}$ and $H_{01}$ describe coupling within the principal unit cell (intra-unit cell) and between the adjacent principal unit cells (inter-unit cell), respectively based on the real space tight binding model given by Eq.\ref{eqn:TBhamiltonian}.

And the real space Hamiltonian can now be written in the desired tridiagonal form:

$$H_\mathrm{ABSNR} =
 \begin{pmatrix}
  H_{AA} & H_{AB} & 0 & 0  & \cdots\\
  H_{AB}^{\dagger} & H_{BB} & H_{BA} & 0 & \\
  0  & H_{BA}^{\dagger} & H_{AA} & H_{AB} &  \\
  0 & 0 & H_{AB}^{\dagger} & H_{BB} & \\
  \vdots & & & & \ddots
 \end{pmatrix},$$

where $H_{AA(BB)}$ and $H_{AB(BA)}$ are intra-unit cell and inter-unit cell $(M \times N)\times (M \times N)$ matrices, respectively.

Furthermore, to understand the effects of the impurity position on the RKKY properties of ABSNRs, we have studied the local density of state (LDOS) of the ABSNRs. Corresponding site-resolved LDOS for site $i-$th, at a given position ${\bf r}$ and energy $E$, is obtained from the imaginary part of the Green's function as $\rho({\bf r},E)=- G^0({\bf r,r},E)/\pi$, calculated using the unperturbed Green’s function matrix as $G^0({\bf r,r},E)=(E-H+i\eta)^{-1}$, where $\eta$ is a positive infinitesimal number.

\subsection{Influence of strain and staggered sublattice potentials on the electronic properties of the ABSNRs}

The controlled introduction of strain into semiconductors, a key strategy for manipulating the magnetic coupling in 2D nanostructures, has a perfect platform for its implementation in the atomically thin materials in both scientific and engineering applications~\cite{Duan17}. Motivated by the search for spintronic materials, a huge number of works have been performed to examine the effectiveness of mechanical strain in modulating the magnetic properties of 2D layered materials~\cite{Pereira09,PhysRevB.76.064120, pereira_tight-binding_2009,ourpaper,Peng20123434, Guinea:gapsgraphene,sharma_effect_2013,Duan17}.
To gain insight on how B$_2$S nanoribbons can be fruitful in the realization of high-performing magnetic devices, fundamental studies on the strain-induced variation of the electronic and magnetic properties of this new material are essential.
In this subsection, the effect of both strain and staggered sublattice potential on the band structure and magnetic exchange interaction is analyzed and discussed.
We first consider an armchair B$_2$S nanoribbon lattice in the $xy$ plane, in the presence of uniaxial strains $\epsilon_x$ and $\epsilon_y$ while a staggered sublattice potential is applied throughout the ABSNR.

Let the $x$-axis be in the direction of the armchair edge of B$_2$S nanoribbon and the $y$-axis in that of the lateral zigzag edge, as shown in Fig.\ref{schem}. Within the context of continuum mechanics and in the linear deformation regime, application of a uniaxial strain will cause the following change of the bond length $r$, in terms of strain components $\epsilon_{x}$ and $\epsilon_{y}$

\begin{eqnarray}
\left(\begin{array}{c}
x'\\
y'
\end{array}\right) & = & \left(\begin{array}{ccc}
1+\epsilon_{x} & \gamma \\
\gamma & 1+\epsilon_{y}
\end{array}\right)\left(\begin{array}{c}
x\\
y
\end{array}\right),
\end{eqnarray}
where ${\bf r}=x {\bf i}+ y {\bf j}$ and ${\bf r'}=x' {\bf i}+ y' {\bf j}$ denote the positions of an atom before and after deformation, respectively.

In the linear deformation regime, an expansion of the norm of $r$ to first order in strains $\epsilon_x$ and $\epsilon_y$ can be
written as $r'\simeq(1+\alpha_x\epsilon_x+\alpha_y\epsilon_y) r,$ where $\alpha_x={({x}/{r})}^2$ and $\alpha_y={({y}/{r})}^2$ are the strain-related geometrical coefficients in the ABSNRs.
According to the Harrison’s formula the transfer integral ($t$) between $s$ and $p$ orbitals scales with the bond length ($r$) as $t\propto\frac{1}{r^{2}}$~\cite{HarrisonWA1999,TangH,J.W.Jiang}.
By invoking the Harrison’s relationship, we get the following geometrical strain effect on the hopping parameter,
\begin{eqnarray}
t & = & t_{0}\left(1-\frac{2}{r}\alpha_{x}\epsilon_{x}-\frac{2}{r}\alpha_{y}\epsilon_{y}\right).
\label{eq_t1}
\end{eqnarray}

One of the fascinating properties of the new families of 2D layered materials is their possibility to use a staggered potential to manipulate their electronic properties. Motivated by this important problem, we examine the effect of staggered sublattice potential on the electronic structure, by breaking the discrete sublattice symmetry of this honeycomb structure. Here, we investigate the band dispersion of the ABSNRs of infinite
length $L$ $(N \rightarrow \infty )$ under the influence of the staggered potential.

\begin{figure}
\includegraphics[width=1\linewidth]{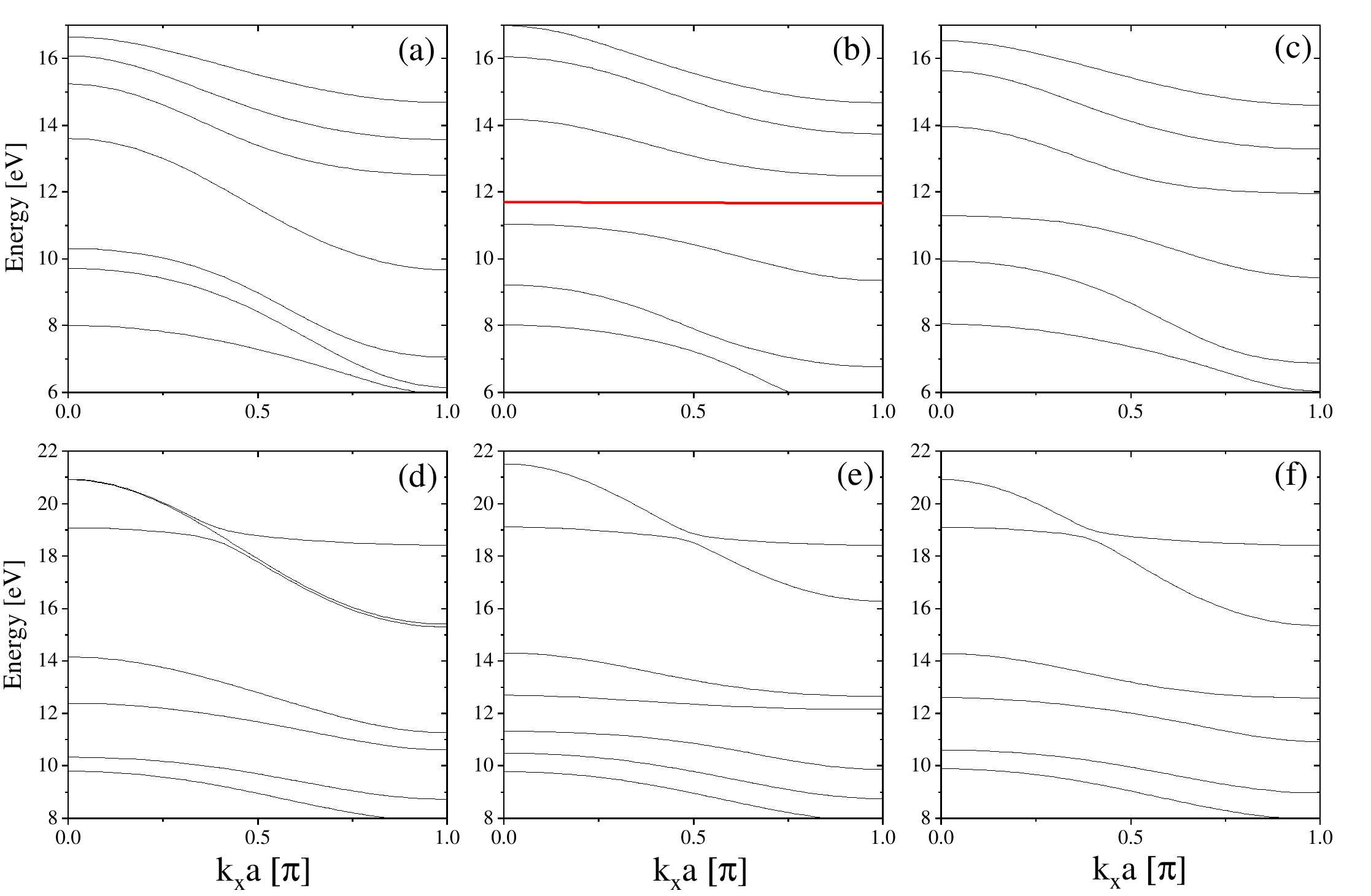}
\caption{(Color online)
The energy band structure for several types of the ABSNRs with infinite lengths with $p=1$ : (a),(d) type 1, (b),(e) type 2, and (c),(f) type 3.
Top panels are for zero staggered potential ($\Delta=0$) and the bottom ones are for nonzero staggered potentials ($\Delta=3$ eV). The quasi-flat band is seen just in the ABSNR of type 2 (red curve).}
\label{DisKp1}
\end{figure}

We first present the calculated electronic band structures of ABSNRs superlattices. The energy band structures of infinite length ABSNRs with width of $M=7$, for different geometry types are plotted in Fig.\ref{DisKp1}. The panels (a),(d) are for ABSNR of type 1, (b),(e) are for type 2, and (c),(f) are for type 3. Top panels are for zero staggered potential ($\Delta=0$) and the bottom ones are for nonzero staggered potentials ($\Delta=3$ eV).
Interestingly, for the ABSNRs of type 2, one can see that a band gap in which a near-midgap band (red curve) completely detached from the bulk bands, is always observed and disappears by introducing staggered sublattice potential term. Indeed, this near-midgap band is shifted and goes away from the flatness mode by applying the staggered potential.
As is known, these near-midgap energy bands are due to the edge states whose wave functions are confined near the ABSNR edges~\cite{H.Zhang,A.CarvalhoEPL,H.Guo14}.

\begin{figure}
\includegraphics[width=1.02\linewidth]{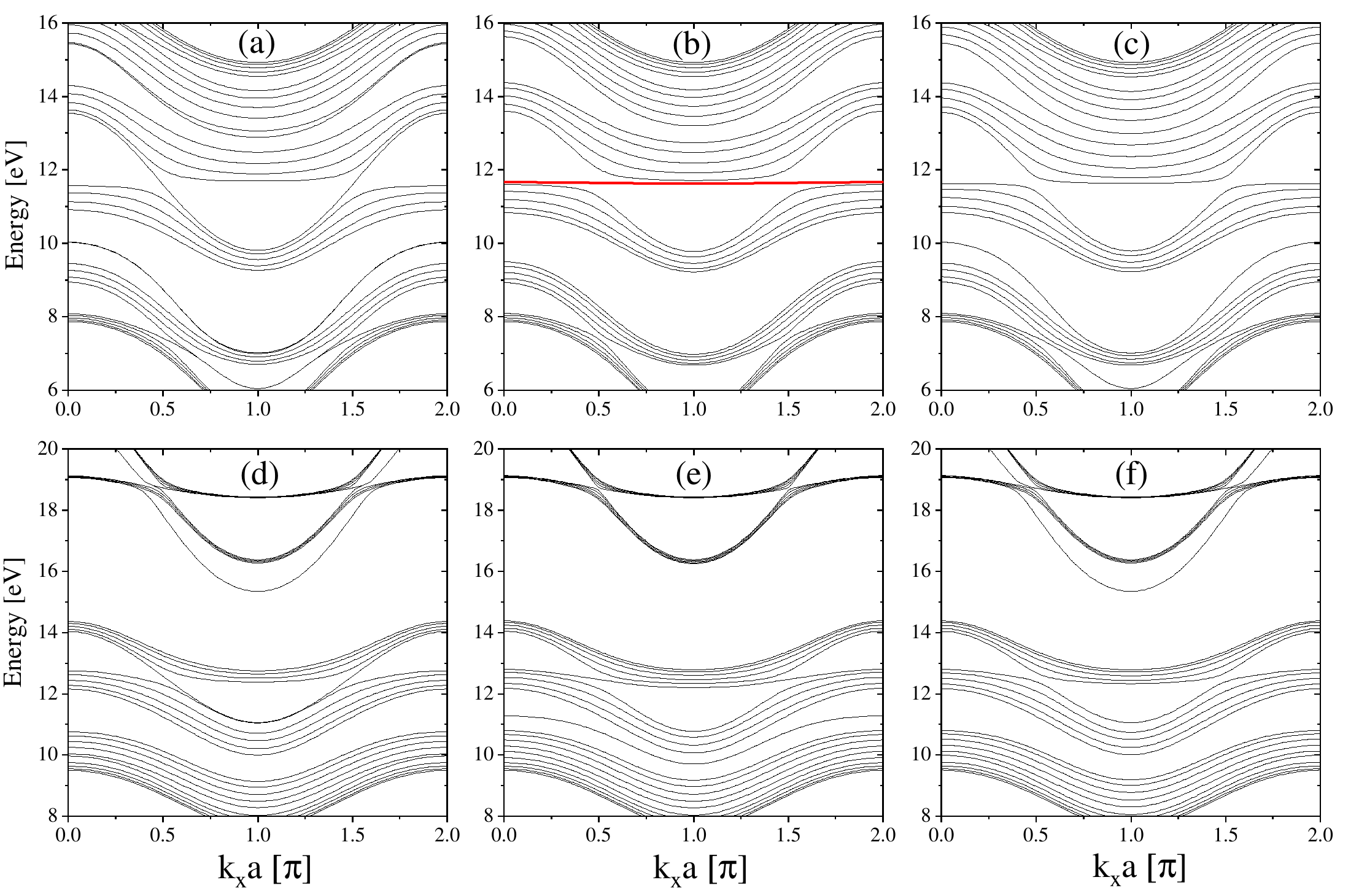}
\caption{(Color online)
The energy band structure for several types of the ABSNRs with infinite lengths with $p=5$ : (a),(d) type 1, (b),(e) type 2, and (c),(f) type 3.
Top panels are for zero staggered potential ($\Delta=0$) and the bottom ones are for nonzero staggered potentials ($\Delta=3$ eV). The quasi-flat band is seen in the ABSNR of type 2 with a red curve.}
\label{DisKP5}
\end{figure}

What the Fig.~\ref{DisKP5} shows is the same as Fig.~\ref{DisKp1} but for ABSNRs with $p=5$. As shown, a large electronic band gap is appeared in the band structure of ABSNRs by applying the staggered potential.
As can be seen, the resulting band structures are completely different at various values of the strength of the staggered potential. The quasi flatband in the ABSNR of type 2, in the absence of staggered potential (panel (b)), is shown with a red color.

\subsection{Quasiflat band tunability in the ABSNRs}
Designing the lattice structures which produce the flat band at Fermi energy has attracted much attention recently because of its potential
applications in nanoelectronics, and magnetoectronics. The presence of flat bands at Fermi energy gives rise to the large density of states and is responsible for the flat band ferromagnetism~\cite{K.Nakada,E.Lieb,K.Kusakabe,M.Fujita96}. There are primarily three ways toward creating flat bands in nanoribbons~\cite{K.Nakada,E.Lieb,K.Kusakabe,M.Fujita96}. The modification of zigzag edge by attaching Klein’s bonds gives rises to the partial flat band in Ggraphene nanoribbons. One of simple ways to obtain the flat bands is given by the nonequality between the sublattice sites in bipartite lattices.
In such lattices, N-degenerated flat bands appear at the Fermi energy. with $N=|N_A-N_B|$, where $N_A$ and $N_B$ are the number of A and B -sublattice sites, respectively~\cite{E.Lieb,J.Fernandez,EzawaPhysicaE,H.Tamura02}.
As suggested by Soleimanikahnoj {\it et al.}, the quasiflat band (midgap-band) modulation provides a platform for pseudospin electronics~\cite{H.Zhang,A.CarvalhoEPL,H.Guo14}, it is interesting to study the band gap and quasiflat band modulation in the ABSNRs.

The aim of this subsection is to elucidate the effect of the both strain and staggered potential on the spectral properties
of the quasiflat edge modes in the ribbon geometry of type 2, specially the formation and tunability of the quasiflat bands in the semiconducting gap in the APNRs.

\begin{figure}
\includegraphics[width=0.92\linewidth]{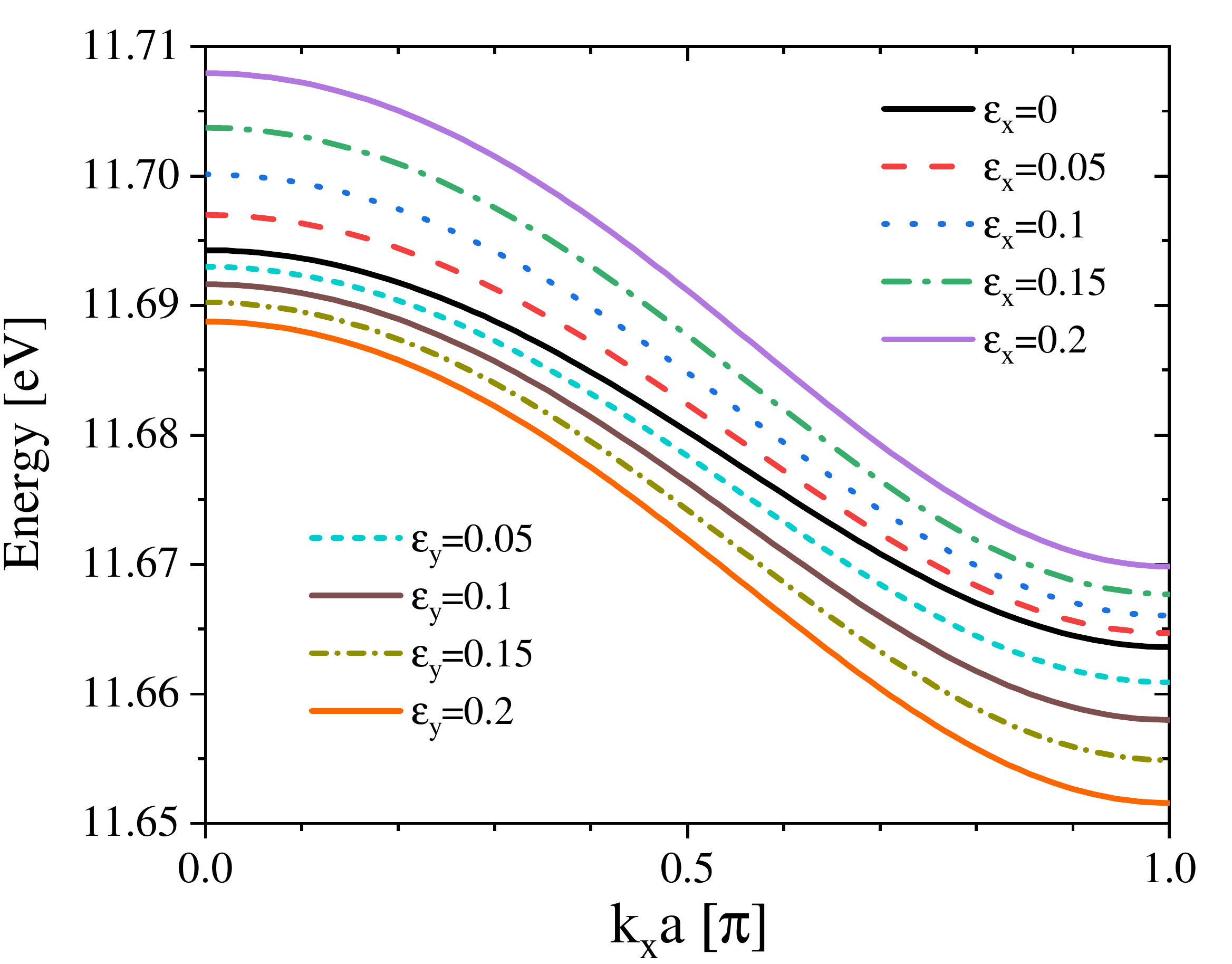}
\caption{(Color online)
Shift of the quasi-flat band for different values of the strains $\varepsilon_x$ and $\varepsilon_y$, for an ABSNR of type 2, with $M=31$. }
\label{figmidgapshift}
\end{figure}

As shown in Fig.\ref{figmidgapshift}, as a key feature, the position of the quasiflat bands in the energy diagram of APNRs can be shifted by applying the in-plane strains $\varepsilon_x$ and $\varepsilon_y$.
Particularly, the quasiflat band energy move up under strains $\varepsilon_x$, while shift down with strains $\varepsilon_y$.

\begin{figure*}
\includegraphics[width=1\linewidth]{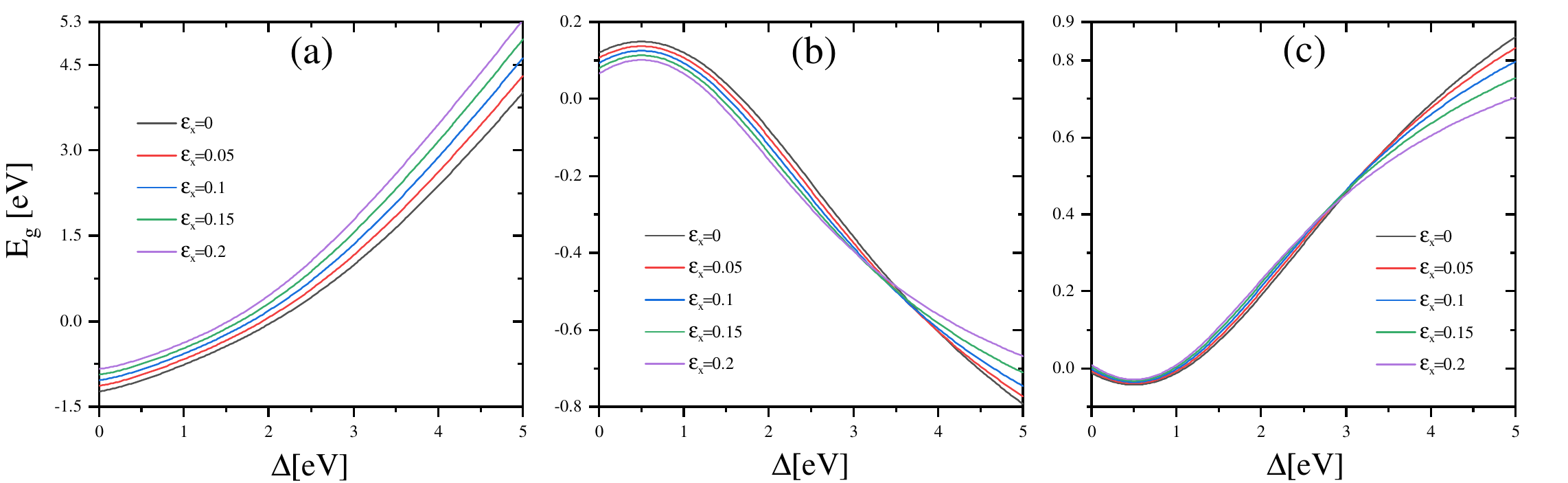}
\caption{(Color online)
Variation of energy band gap of ABSNRs with $p=5$ as a function of sublattice staggered potential for different values of strain $\varepsilon_x$.}
\label{figEg}
\end{figure*}

To achieve a superior performance of the ABSNRs in optoelectronic devices based on B$_2$S nanoribbons, a feasible band gap modification is crucial for nanoribbons.
To reveal the staggered potential dependence of the band gap of ABSNRs, we have calculated the band gaps for all three types of ABSNR superlattices as a function of the applied staggered potential $\Delta$ for different values of strain $\varepsilon_x$ (see Fig.\ref{figEg}).
It is visible that the band gap energy of ABSNRs depends strongly upon the applied staggered potential $\Delta$ and thus one can engineer the electronic properties of the ABSNRs via tuning the external staggered potential.
From the Fig.\ref{figEg} a complete and fully reversible semiconductor (or insulator) to metal transition has been observed via tuning the external staggered potential, which can be easily realized experimentally. It should be emphasized that a negative energy gap corresponds to a metallic state and a positive energy gap corresponds to a semiconductor or insulator electronic state, depending on the energy gap values.

As we have demonstrated, depending on the applied staggered potential in various strain configurations, one may have ABSNRs with favor electronic structure, namely, semiconductor, insulator or the metallic state.

We elaborated more on the potential tunability below and show how to get highly improved flat bands with perfect flatness.

\begin{figure}
\includegraphics[width=0.9\linewidth]{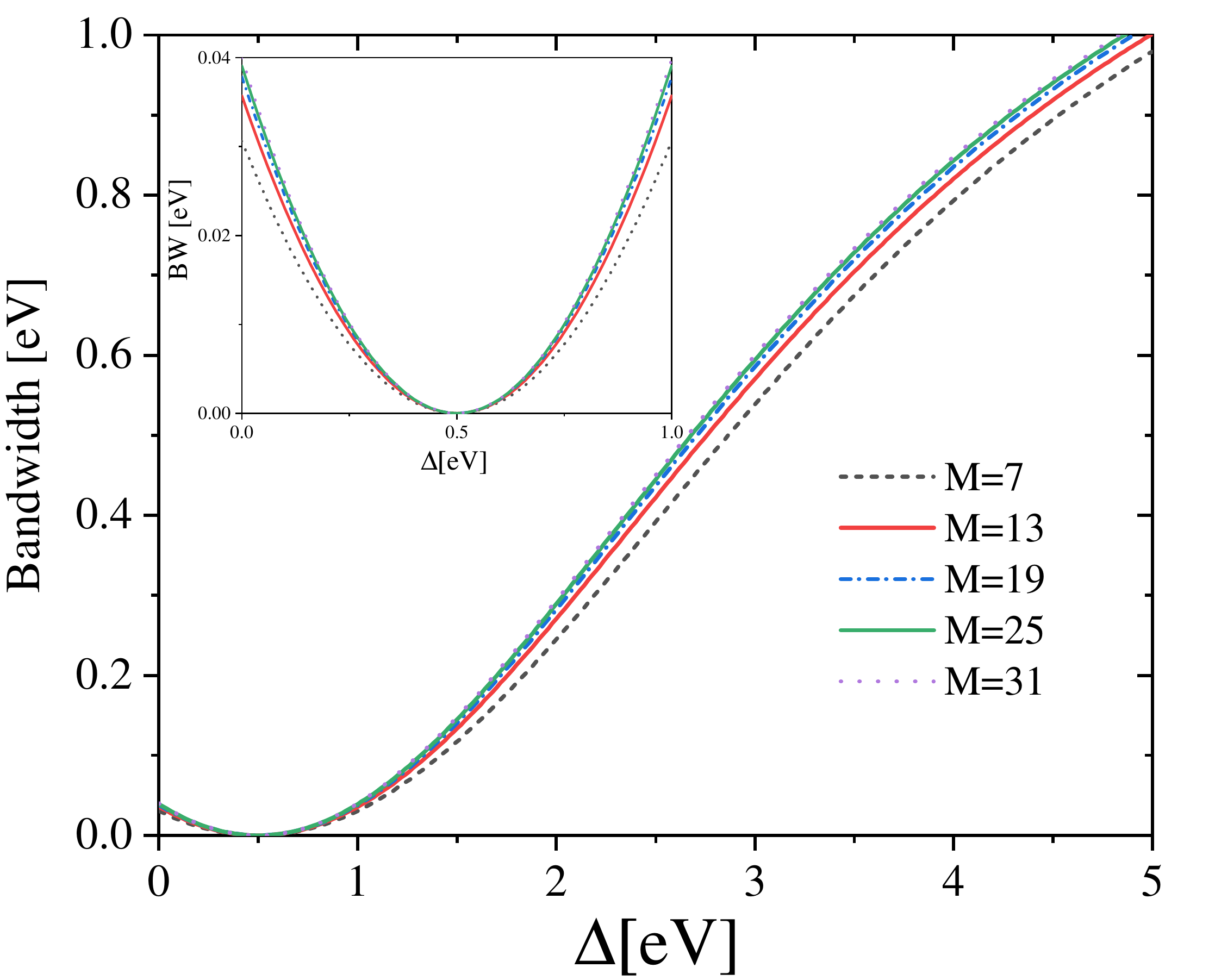}
\caption{(Color online)
The effect of the staggered sublattice potential $\Delta$ on the flatband bandwidth (FBBW) for ABSNRs with different widths.}
\label{FigMidg}
\end{figure}

Figure \ref{FigMidg} presents the flat band bandwidth versus the applied staggered potential for ABSNRs with different widths. It is clear that, the response of the midgap bands to the applied staggered potential depends on the width of the ribbon. It is worthwhile to note that the bandwidth is generally defined as the energy difference between the upper and lower band edges.
It is important to note that at a critical staggered potential ($\Delta_c=0.5$ eV), for ABSNRs with any width, the quasi-flatband changes to a perfect flatband. Moreover, at a fixed staggered potential the midgap bandwidth (MBW) decreases with increasing the width of the ABSNR. Interestingly, the graph of bandwidth for ABSNRs with any width is symmetric with respect to the critical staggered potential (at the interval [0-1] eV), as shown in the inset. For staggered potentials greater than the critical staggered potential ($\Delta>\Delta_c$), the bandwidth of the quasi flat bands monotonically increases with increasing the staggered potential. For staggered potentials smaller than the critical staggered potential ($\Delta<\Delta_c$), the trend in reverse {\i.e.}, the bandwidth of the bands decreases with increasing the sublattice staggered potentials.
As shown, the critical staggered potential (the potential at which semiconductor (or insulator)-to-metal transition occurs) is different for different types of the ABSNRs and changes with strain.

\subsection{Numerical results for the RKKY interaction in zigzag B$_2$S nanoribbons }\label{sec:Numer-R}
In this section, we present in the following our main numerical results for the numerical evaluation of the RKKY exchange (Eq.\ref{chiE}) in the armchair B$_2$S nanoribbons, based on the tight-binding model (Eq.\ref{eqn:TBhamiltonian}). For simplicity, all obtained data for the RKKY interaction are multiplied by $10^3$.

\begin{centering}
\begin{figure*}
\includegraphics[width=1.0\linewidth]{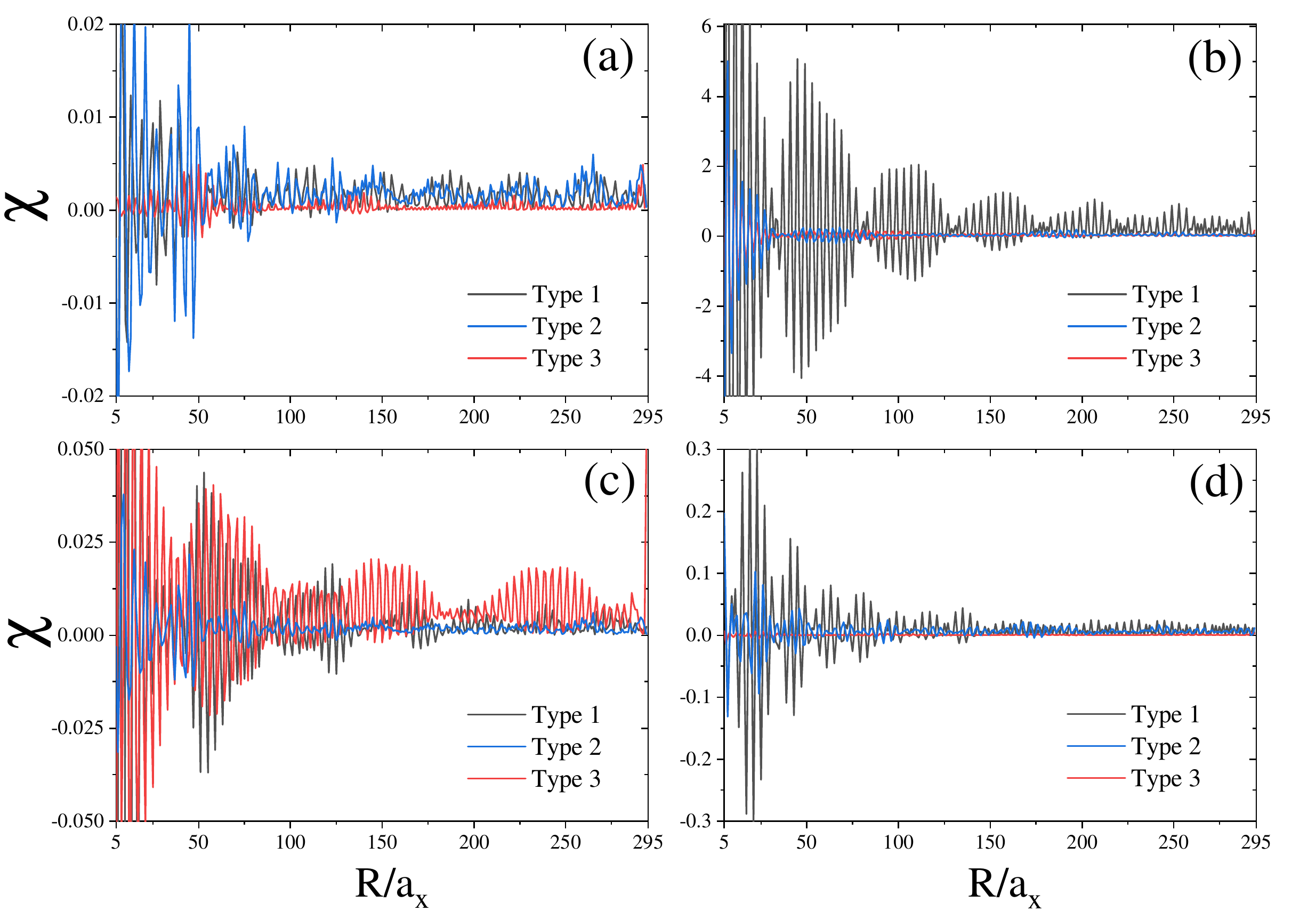}
\caption{(Color online) The variation of $\chi$ versus distance between two impurities, for doped ABSNRs with $N=300, p=2$, for different types of ABSNRs with $E_F=2$ eV. The details of the panels are as follows:
(a) Both the impurities are located at the same edge, the first impurity at the edge site with coordinate $(5,1)$ and the second one at lattice sites $(n,1)$.
(b) Both the impurities are situated in the interior region of the ABSNR, the first impurity at lattice site $(5,7)$ and the second impurity at lattice points $(n,7)$,
(c) One impurity is at the edge site $(5,1)$ and the other one is moved interior of the ABSNR along the line $n=7$ at the lattice sites $(n,7)$,
and (d) The impurities are located at the opposite edge sites (interedge magnetic coupling) with coordinates $(150,1)$, $(150,13)$ for the ABSNRs of types 1 and 2 and $(150,1)$, $(150,12)$ for the ABSNR of types 3.}
\label{chi_R0}
\end{figure*}
\end{centering}

Figure \ref{chi_R0} shows the effective exchange interaction for doped ABSNRs ($E_F=2$ eV ) with $N=300$ and $p=2$ as a function of distance between the impurities for possible impurity configurations for different types of ABSNRs. The details of the panels are as follows:
(a) Both the impurities are located at the same edge, the first impurity at the edge site with coordinate $(5,1)$ and the second one at lattice sites $(n,1)$.
(b) Both the impurities are situated in the interior region of the ABSNR, the first impurity at lattice site $(5,7)$ and the second impurity at lattice points $(n,7)$,
(c) One impurity is at the edge site $(5,1)$ and the other one is moved interior of the ABSNR along the line $n=7$ at the lattice sites $(n,7)$,
and (d) The impurities are located at the opposite edge sites (interedge magnetic coupling) with coordinates $(150,1)$ and $(150,13)$ for the types 1 and 2 and $(150,1)$ and $(150,12)$ for the ABSNR of types 3.

It is worth pointing out that for small distance between the impurities, the impurities interact very strongly with each other, but
then rapidly decay with R until its flattens out to a constant value.
As a result, a beating pattern of oscillations of the RKKY interaction occurs for all types of ABSNR, for the doped systems.
It is clear that the edge structure of the ABSNRs has a strong effect on the RKKY coupling.
The edge-geometry contributions to the RKKY interaction were found to be more important for the geometry with both impurity
spins are situated in the interior of the ABSNR (panel (b)), because in this configuration the RKKY interaction is very strong for edge-geometry of type 1 in comparison to the other two geometries.
For the case when both spins are inside the ABSNRs (the panel (b)) the result is quite different, because in this situation the RKKY interaction is at least one order of magnitude greater than the other configurations.

\begin{centering}
\begin{figure*}
\includegraphics[width=1.0\linewidth]{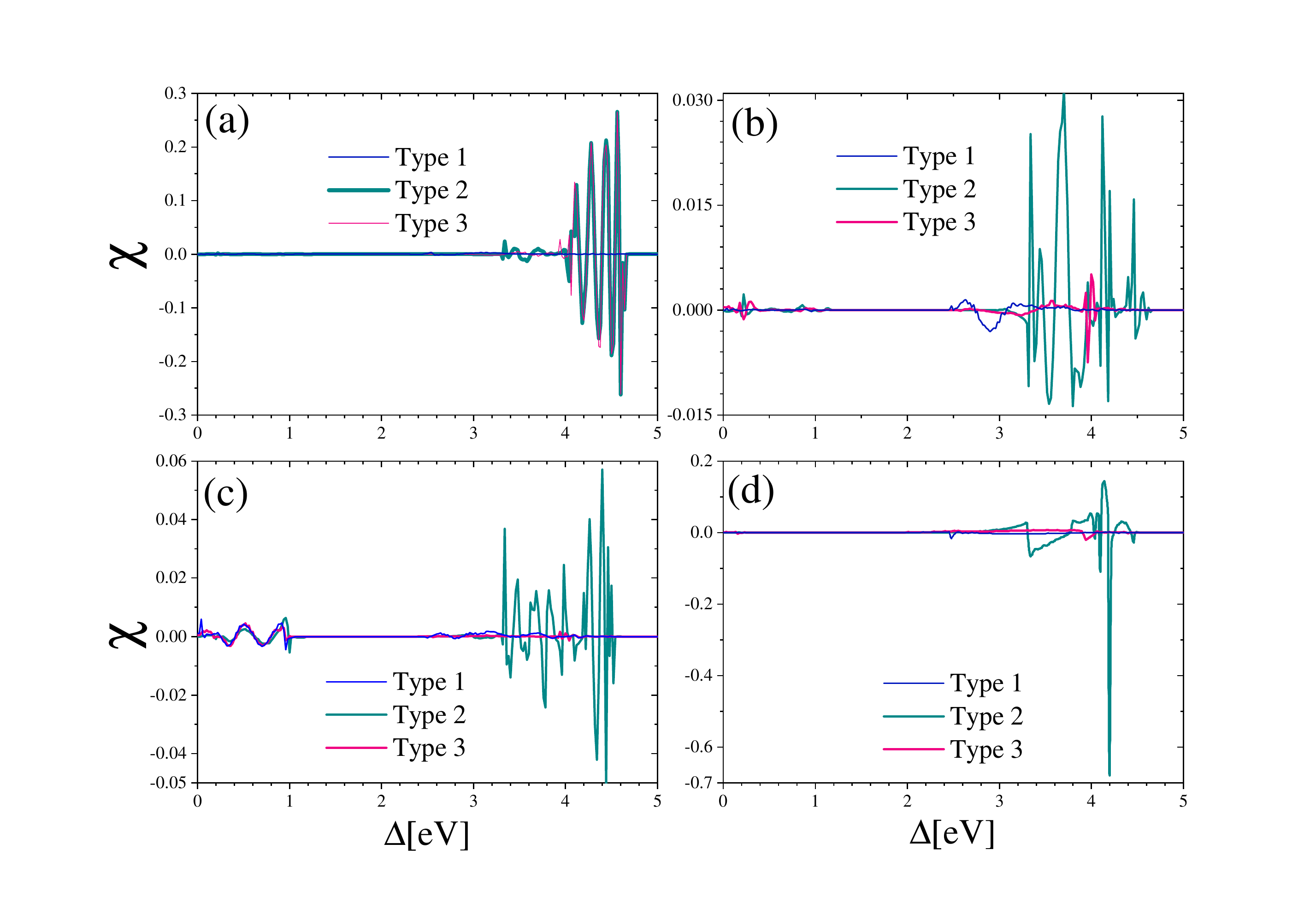}
\caption{(Color online) The variation of $\chi$ with applied staggered potential for different types of the ABSNRs with $p=2, N=300$, for a fixed spatial distance between impurities. (a) Both the impurities are located on the same edge at the edge sites with coordinates $(145,1)$ and $(155,1)$
(b) Both the impurities are located in the interior of the ABSNR, on lattice points with coordinates $(145,7)$ and $(155,7)$,
(c) One of the impurities is located on the edge site $(145,1)$ and the second one is on the lattice site $(155,7)$, and
(d) The impurities are located on the opposite edge sites (interedge magnetic coupling) with coordinates $(150,1)$ and $(150,13)$.}
\label{chidelta}
\end{figure*}
\end{centering}

The staggered potential dependence of the RKKY interaction is shown in Fig.\ref{chidelta}, where different distance configurations between two impurities are considered. (a) Both the impurities are located on the same edge at the edge sites with coordinates $(145,1)$ and $(155,1)$
(b) Both the impurities are located in the interior of the ABSNR, on lattice points with coordinates $(145,7)$ and $(155,7)$,
(c) One of the impurities is located on the edge site $(145,1)$ and the second one is on the lattice site $(155,7)$, and
(d) The impurities are located on the opposite edge sites with coordinates $(150,1)$ and $(150,13)$.

It is shown that the RKKY interaction has an oscillating behaviour in terms of the applied staggered potentials, such that for two magnetic adatoms randomly distributed on the surface of an ABSNR the staggered potential can reverse the RKKY from antiferromagnetism to ferromagnetism and vice versa.
Importantly, the various edge geometries of ABSNRs show tunability in magnetic RKKY coupling on the application of external staggered potentials, strain. This proves to be an alternative approach to tuning the impurity interactions in ABSNRs.

\begin{figure}
\includegraphics[width=1.02\linewidth]{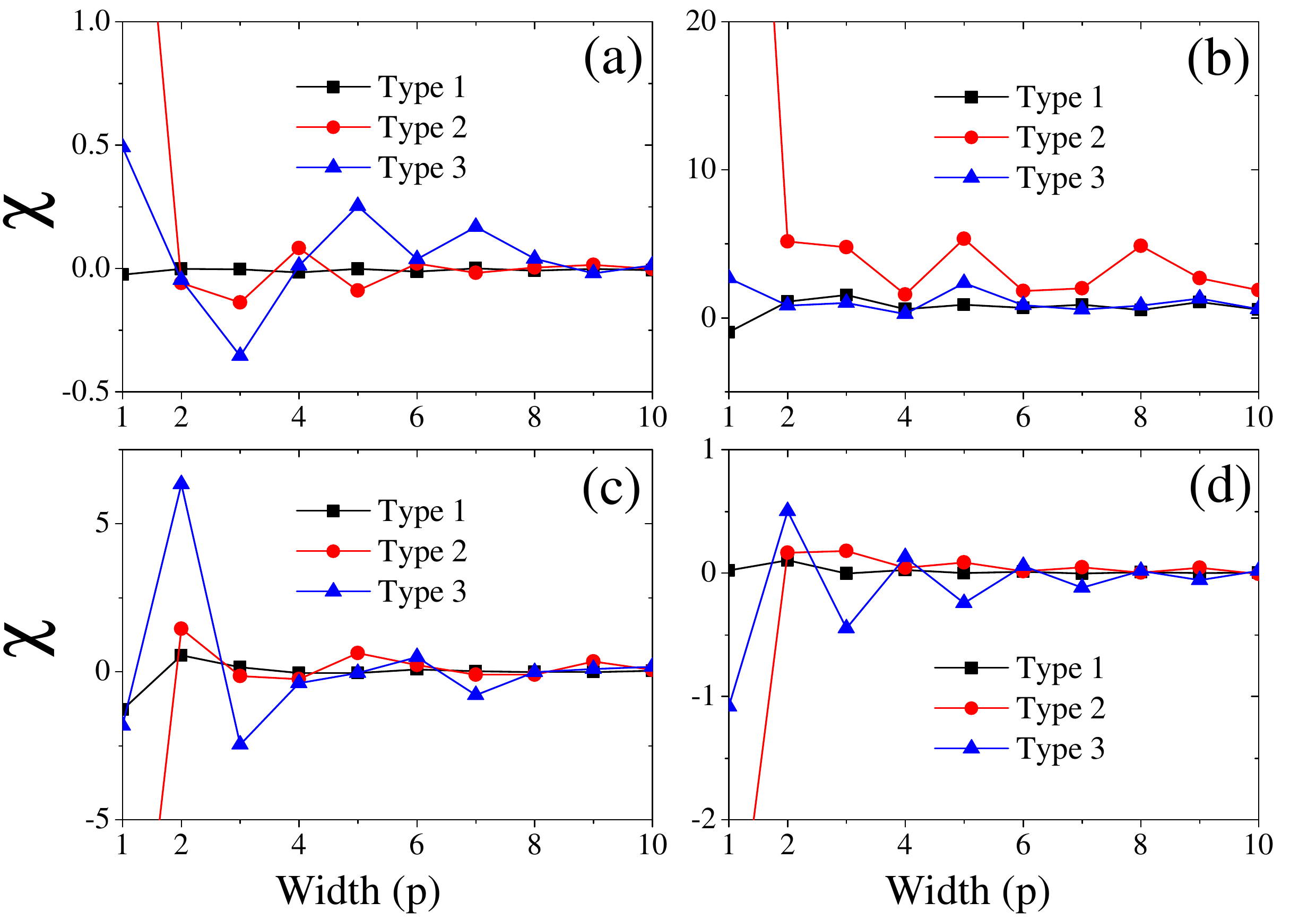}
\caption{(Color online) The variation of $\chi$ as a function of the system width for ABSNRs with $N=150$, in the absence of both strain and staggered potential.
(a) Both the impurities are located on a same edge, (b) Both the impurities are located on the interior of the ABSNR, (c) One of the impurities is located on an edge site and the second one is on a bulk lattice site and (d) The impurities are located on the opposite edge sites.}
\label{chi_Width}
\end{figure}

We further investigate the effect of the nanoribbon's width and edge geometry on the RKKY exchange coupling in ABSNRs (see Fig.~\ref{schem}).
Figure \ref{chi_Width} shows the dependence of the RKKY coupling on the ribbon width for all three types of ABSNRs.
The RKKY interaction has an oscillatory behavior in terms of the width of the ribbon. One can find from these figures that, with an increase
in width, the exchange couplings drop at first and then their oscillating amplitudes decay with increasing the width of the ABSNRs finally approach converged value (almost zero).
Such an oscillatory behavior versus the ribbon's width for graphene has been reported previously~\cite{N.Gorjizadeh}.
We observe that for the case that both of two impurities are situated within the interior of the nanoribbon (panel b) the magnetic coupling of ABSNRs with a finite width is always ferromagnetism that is very robust against the impurity movement.

\begin{figure}
\includegraphics[width=0.9\linewidth]{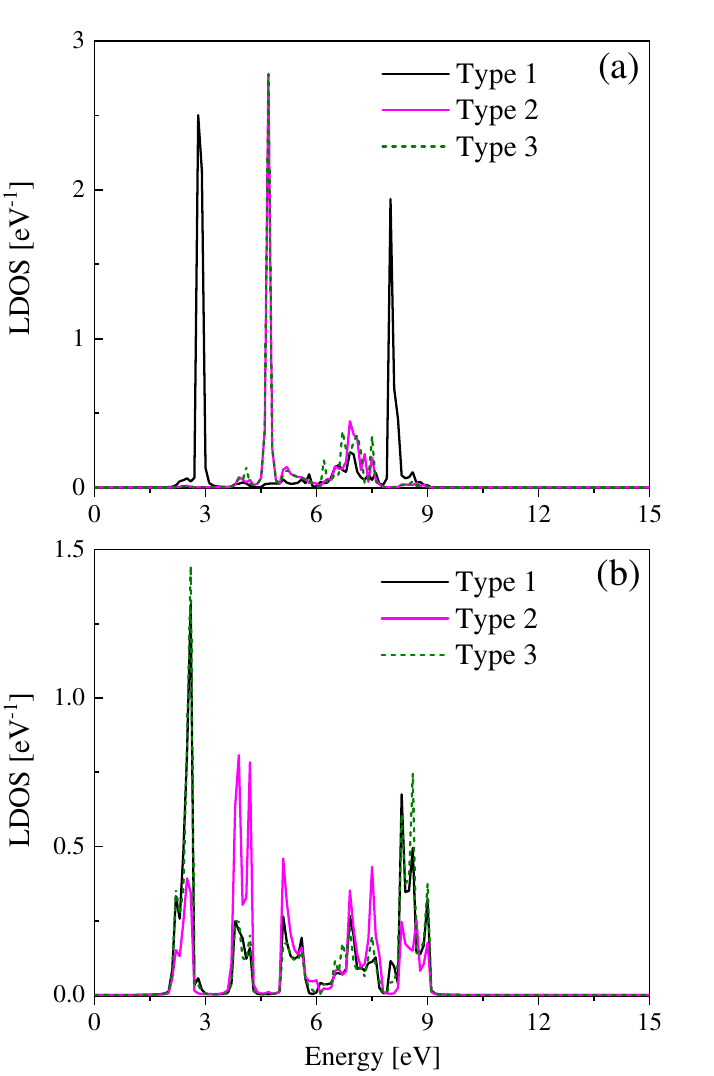}
\caption{(Color online)
Calculated local density of states for an ABSNR with $N=300$, $p=2$, for both edge and bulk sites: (a) an edge lattice site with coordinate $(150,1)$,  (b) a bulk lattice site with coordinate $(150,7)$.
}
\label{LDOS}
\end{figure}

As is known, understanding the sublattice-dependent of local density of states (LDOS) is essential to assess the configuration-dependent exchange interaction. To do so, it is necessary to obtain the diagonal components of the unperturbed Green’s function matrix ($ G({\bf r,r},E)$), for a lattice site at position $ {\bf r}$ and energy E.
Fig.~\ref{LDOS} illustrates the LDOS for an ABSNR with $N=300$, $p=2$, for both edge and bulk sites: (a) an edge lattice site with coordinate $(150,1)$  and (b) a bulk lattice site with coordinate $(150,7)$.

Clearly for ABSNRs of type 2 and type 3, there is a high LDOS peak in the edge sublattice, for energies around $E\sim4.7$ eV (panel (a)).
Here also two peaks at different energies around $E\sim2.8$ and $E\sim 8$ eV appear for ABSNR of type 1.
On the contrary, for a bulk site there is a high LDOS peak for energies around $E\sim2.7$ eV for ABSNRs of type 1 and type 3 (panel (b)).

\section{summary}\label{sec:summary}.

To summarize, in this work, we numerically investigate the RKKY exchange coupling between two magnetic impurities located on an armchair B$_2$S nanoribbon, a new anisotropic Dirac cone material, as a strained graphene.

In the first part of our study, employing a tight-binding approach, we investigate the electronic properties of armchair-terminated B$_2$S nanoribbons in the presence of both strain and staggered sublattice potential.
It is found that armchair B$_2$S nanoribbons (ABSNRs) show different electronic and magnetic behaviors due to different edge morphologies.
The band gap energy of ABSNRs depends strongly upon the applied staggered potential $\Delta$ and thus one can engineer the electronic properties of the ABSNRs via tuning the external staggered potential.
A complete and fully reversible semiconductor (or insulator) to metal transition has been observed via tuning the external staggered potential, which can be easily realized experimentally.
Interestingly, for the ABSNRs belong to the family $M=6p$, with $M$ the width of the ABSNR and $p$ an integer number, one can see that a band gap, in which a quasi-flatband completely detached from the bulk bands is always observed.
As a key feature, the position of the quasi-flatbands in the energy diagram of ABSNRs can be shifted by applying the in-plane strains $\varepsilon_x$ and $\varepsilon_y$.
At a critical staggered potential ($\Delta_c \sim0.5$ eV), for ABSNRs with any width, the quasi-flatband changes to a perfect flatband.

Then, within the tight-binding model we exploit the Green's function formalism, to reveal how the RKKY interaction between the impurities placed on a ABSNR is affected by mechanical strain, in the presence of a staggering potential.
In particular, the effects of ribbon width,strain and staggered sublattice potential on the behavior of RKKY interaction have been investigated.
For impurities at fixed values distance, the increase of applied staggered potential leads to higher values of exchange coupling.
It is shown that the RKKY interaction has an oscillating behaviour in terms of the applied staggered potentials, such that for two magnetic adatoms randomly distributed on the surface of an ABSNR the staggered potential can reverse the RKKY from antiferromagnetism to ferromagnetism and vice versa. The RKKY interaction has an oscillatory behavior in terms of the width of the ribbon.
It is shown that the magnetic interactions between adsorbed magnetic impurities in ABSNRs can be manipulated by careful engineering of external staggered potential.
Therefore, the ABSNRs would be expected to be a very promising candidate for spintronics and pseudospin electronics devices based on ABSNRs.
\section*{References}

\end{document}